\documentclass[12pt]{iopart}

\usepackage{amssymb}
\usepackage{iopams}
\usepackage{setstack}
\usepackage{graphicx}

\begin{document}

\title[Finding the optimum activation energy ... Chaudhury et al.]
{Finding the optimum activation energy in DNA breathing dynamics:
A Simulated Annealing approach}

\author{Pinaki Chaudhury$^1$, Ralf Metzler$^2$ and Suman K Banik$^3$}

\address{$^1$Department of Chemistry, University of Calcutta, 92 A P C
Road, Kolkata 700 009, India.}
\address{$^2$Department of Physics, Technical University of Munich,
D-85747 Garching, Germany.}
\address{$^3$Department of Chemistry, Bose Institute, 93/1 A P C Road,
Kolkata 700 009, India.}

\ead{pinakc@rediffmail.com, metz@ph.tum.de, skbanik@bic.boseinst.ernet.in}

\date{\today}

\begin{abstract}
We demonstrate how the stochastic global optimization scheme of Simulated 
Annealing can be used to evaluate optimum parameters in the problem of DNA 
breathing dynamics. The breathing dynamics is followed in accordance with the 
stochastic Gillespie scheme with the denaturation zones in double stranded DNA 
studied as a single molecule time series. Simulated Annealing is used to find 
the optimum value of the activation energy for which the equilibrium bubble
size distribution matches with a given value. It is demonstrated that the
method overcomes even large noise in the input surrogate data.
\end{abstract}



\maketitle

\section{Introduction}

Optimization techniques have been successfully used across disciplines
\cite{Fogar,Pulay,Sch,Head,Wales,Bacelo,Berne}. In general, a given problem
is cast in a manner such that finding the extremum point of a functional in
some search space renders the desired solution. For illustration, consider
the non-trivial problem of finding the global minimum in a rugged potential
energy surface. In this problem one starts from any arbitrary point, and then
moves on the search space following certain criteria (one could be that a move
is accepted if the gradient norm decreases in a given move) and converge on
a point for which the gradient norm is zero. To verify whether the obtained
point is indeed a minimum, or not, one needs to check if the eigenvalues of
the associated Hessian matrix at that point are all positive, or not. In
general, for any problem, in which a finite number of optimum values of
parameters are sought after, one can write down an adequate functional and
extremize it by following the above procedure.

However an optimization procedure whose mode of action is solely guided by
minimizing the local gradient norm (one which is deterministic) may face
difficulties, especially in search spaces with multiple minima. If the
obtained minimum is a local one, there is no way of escaping from it again
and moving towards the global minimum. Here the concept of true global
optimizers, whose search is not solely driven by a local gradient, or one
which is generically stochastic in nature, is needed. Simulated Annealing (SA)
is such a global optimizer and has been very successfully used since it
was originally introduced by Kirkpatrick et al. \cite{ksk}. In
particular, in these references SA was used to tackle the famed traveling
salesman problem (a so-called NP--nondeterministic polynomial time problem).
Due to its successful application a large amount of literature on SA has been
published across fields, see, for instance,
Refs.~\cite{Car,Nandy,PDutta,Ming,Zuck}.

SA is a search technique which borrows its concepts from the physical process
of annealing. If one is trying to produce a good alloy, a molten state of the
metallic mixtures is prepared at a very high temperature (starting annealing
temperature $T_{at}$) and then gradually cooled. If the rate of cooling 
(\emph{annealing schedule}) is slow enough the system will solidify at the
minimum (stable) thermodynamic state. SA exactly follows this principle.
The search space is sampled initially at a high temperature ($T_{at}$), and
then gradually at lower values of $T_{at}$ determined by a preassigned
schedule. The temperature $T_{at}$ controls the strength of thermal
fluctuations, so that even if the system is trapped in a local minimum, a 
high enough $T_{at}$ can take it out of the attractive basin and the search
can carry on to locate deeper minima, and eventually the global one. In the
limit of very small $T_{at}$ or if the search is carried on for a sufficient
length of time, the global minimum will be unequivocally found. Operational
details of this procedure which will be presented below.

In this communication we employ SA to study the dynamics of DNA denaturation
bubbles. We show that SA finds the optimum value of the activation energy
for base pair breaking and therefore reproduces correctly the bubble free
energy landscape. Moreover we demonstrate that SA performs well even in the
presence of strong noise superimposed on the surrogate input data. In the
optimization procedure we start with an arbitrary value of the activation
energy and allow SA to improve it until the bubble size distribution matches
a given distribution. We specifically use SA in this problem as we believe it
is well suited to analyze experimental data on DNA bubble dynamics.

The article is organized as follows. In the following we first discuss the
methodology of SA in detail and connect it to the optimization of the
activation energy required for DNA base pair opening. Then the simulation
results are presented and discussed; and finally the article is concluded.

\section{DNA bubble dynamics and Simulated Annealing}
  
\begin{figure}
\includegraphics[width=6cm]{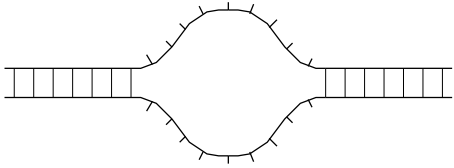}
\caption{DNA denaturation bubble. The bubble consists of single-stranded DNA
that is embedded in still intact double-strand. At the interface between
bubble and double-strand base pairs break or reanneal stochastically, giving
rise to the bubble dynamics.}
\label{bubble}
\end{figure}

Double-stranded DNA can locally denature, i.e., the base pairs usually
forming the double-helix break \cite{dp}. The result is a DNA bubble
consisting of single-stranded DNA, see Fig.~\ref{bubble}. Driven by
thermal fluctuations this bubble changes its size by breaking or
reannealing of base pairs at the two interfaces between bubble and
intact double-strand \cite{dp}. The resulting DNA breathing, or DNA
bubble dynamics therefore is a stochastic process \cite{hame,skb,tobias}.
To follow the time series of the kinetics one needs to resort to a
formulation that intrinsically includes the fluctuations in an efficient
way. It was shown in Ref.~\cite{skb} that the Gillespie algorithm \cite{dtg},
in which the multi-step rate equation is written down as a single master
equation, provides time series and equilibrium quantities of bubble breathing
dynamics.

\begin{figure}
\includegraphics[width=0.7\columnwidth,angle=-90]{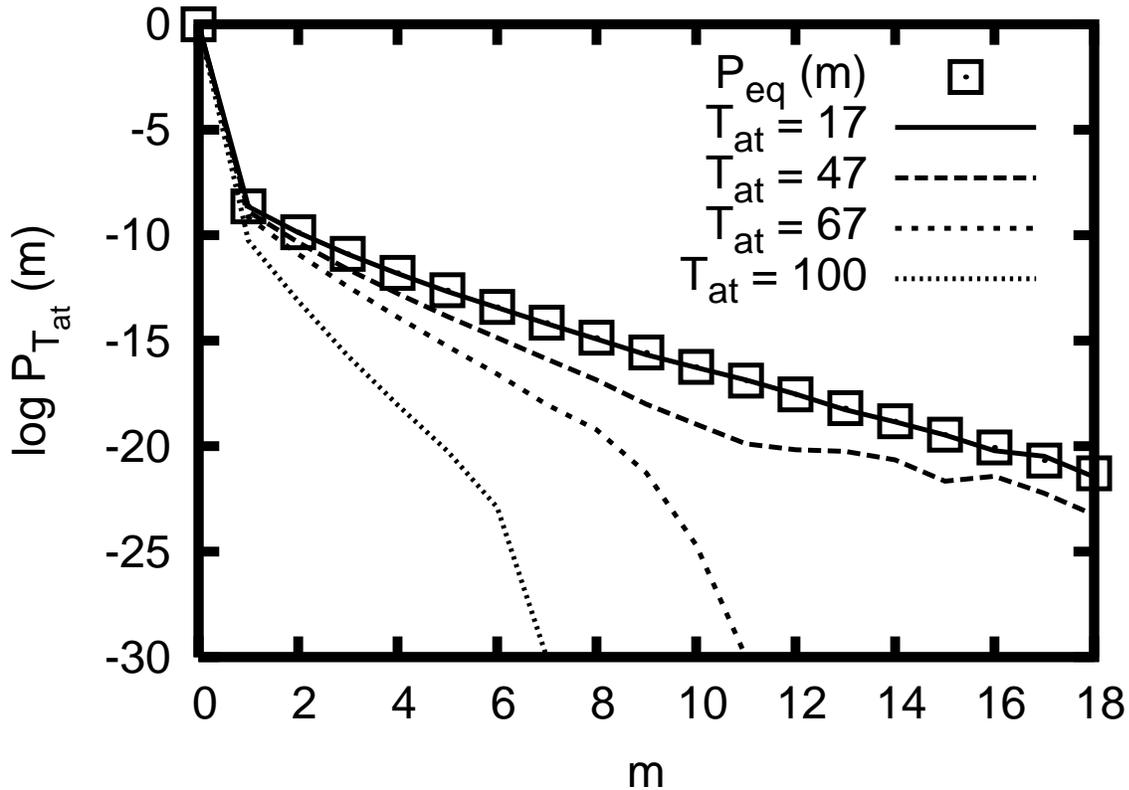}
\caption{Plot of the equilibrium probability $P_{\mathrm{eq}}(m)$ to find a
bubble of $m$ broken base pairs for a bubble domain of maximum size $m=18$.
The parameters for the open boxes 
are $\sigma_0=10^{-3}$ and $u=0.6$. We also show the SA-estimates
$P_{T_{at}}(m)$ at various annealing temperatures $T_{at}$. The plot
for $T_{at}= 17$ matches quite well with the preset surrogate equilibrium 
distribution, $P_{eq}(m)$.}
\label{prob_dist}
\end{figure}

A denaturation bubble opens (nucleates) with an initiation free energy factor
$\sigma_0=\exp(-E_s/k_BT)\approx10^{-5}\ldots10^{-3}$ at room temperature
\cite{blake}. Once initiated, additional base pairs break with a free
energy $E(T)$, with associated statistical weight $u=\exp(-E(T)/k_BT)$.
This free energy $E(T)$ is composed of the free energies for hydrogen bonding
and base stacking \cite{krueger}. In general the value of $E(T)$ will depend
on the type of nearest neighbor base pairs (AT or GC) \cite{krueger},
however, here we consider a homopolymer DNA in which all $E(T)$ are equal.
Thermal melting of the DNA double-strand defining the melting temperature
$T_m$ occurs when $E(T_m)=0$. Below the melting temperature the fully closed
(no bubble) state will always be favorable. However fluctuations may initiate
a transient bubble. The ensuing opening/closing dynamics of a single DNA bubble
can in fact be monitored in fluorescence correlation experiments \cite{oleg}.
At equilibrium the probability distribution to find a bubble of size $m$
broken base pairs is plotted in Fig.~\ref{prob_dist} with logarithmic ordinate.
The initial jump corresponds to the initiation free energy factor $\sigma_0$.
Subsequently one observes a straight decay corresponding to the investment of
the free energy $E(T)$ per broken base pair.

In this work we wish to employ a stochastic optimization search for the 
correct determination of $u$ at the particular temperature in which the system
is studied. For this we have a theoretical probability distribution profile
for the occupation at various DNA sites \cite{skb} with 
which we will compare our distribution profiles for various $u$ values
determined for different annealing temperatures given by the annealing
protocol. SA will guide our search for the optimum $u$ such that we arrive
at the theoretical distribution profile in as few iterations as possible.

In the SA algorithm thermal fluctuations are used to cross barriers in the
underlying search landscape such that optimum parameters can be found even
in a rugged surface. The simulation is started at a sufficiently high
annealing temperature $T_{at}$. This makes nearly all moves acceptable as the 
criterion for accepting or rejecting a move is determined by the Metropolis
criterion. In our case the objective function (often called \emph{cost
function\/} in the SA language) which is being minimized is the sum of the
squares of the difference of the occupation probabilities at the various sites 
\begin{eqnarray}\label{eq1}
\mathrm{cost} = \sum_{i=1}^{m} ( P_{eq}(m)- P_{T_{at}}(m))^2 ,
\end{eqnarray}
where $P_{eq}(m)$ denotes the preset surrogate probability distribution to
have $m$ broken base pairs, and $P_{T_{at}}(m)$ is the distribution obtained
from SA at temperature $T_{at}$. If, on going from one step to the other in
the annealing protocol defined below, the magnitude of the cost function
decreases, this move is accepted. If it increases, the move is not discarded
immediately. We subject it to the Metropolis test \cite{Metrop}: The quantity 
\begin{equation}
\Delta=\mathrm{cost}_i-\mathrm{cost}_{i-1}
\end{equation}
is computed where $\mathrm{cost}_i$ is the magnitude of the cost in the
present step and ${\rm cost}_{i-1}$
is the value at the step before. As we observe an increase in the cost
function, the quantity $\Delta$ is positive. The probability of accepting this
move is then determined by evaluating the function
\begin{eqnarray}
\label{eq2}
F=\exp(-\Delta/T_{at}).
\end{eqnarray}

\noindent By definition $F$ ranges between 0 and 1. For each evaluation of $F$ we
invoke a random number $Rn$ between 0 and 1. If $F$ is greater than $Rn$
we accept the move, following the idea that the cost increase corresponding
to the ``Boltzmann factor'' $F$ is smaller than the intensity of the
fluctuation corresponding to the random number $Rn$. If $F<Rn$ the move
is rejected. Thus at high $T_{at}$, $F$ will be close to 1 and most moves
will be accepted, and a greater region of search space will be sampled.
As the simulation proceeds, $T_{at}$ is lowered by following the annealing
schedule,
as the optimization proceeds. This decrease in $T_{at}$ can be a simple
decrease by a certain factor or an exponential decrease depending
on the nature of the problem.
Once on the right path towards the global minimum we need not
search the entire space and concentrate on a small region which will
funnel the search specifically to the global minimum. As $T_{at}$ is
lowered less and less moves pass the Metropolis test. More and
more of the accepted moves correspond to a decrease in the cost function
and we move towards the global minimum. In the current setup it is the
value of the base pair breaking free energy factor $u$ for which $F$
is minimized.

\section{Results and Discussion}

\begin{figure}
\includegraphics[width=0.7\columnwidth,angle=-90]{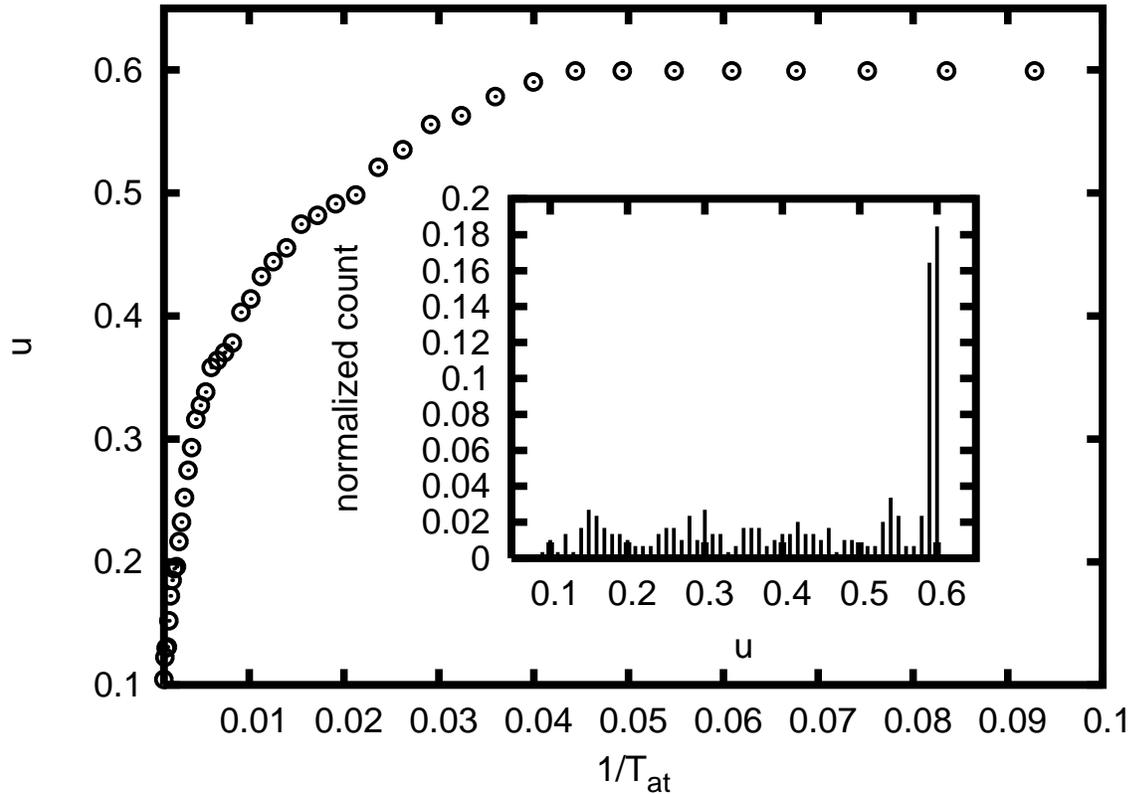}
\caption{Plot of the SA estimate of the statistical weight factor $u$
versus the annealing temperature $1/{T_{at}}$. Inset: Histogram showing the
Metropolis sampling of instantaneous fluctuations in $u$ for the entire range
of annealing temperature performed during the 
optimization process. The main figure plots the best $u$ at a given
temperature step obtained from the fluctuating data shown in the histogram.}
\label{uprofile}
\end{figure}

We focus on the evolution of a few quantities important to characterize the
denaturation bubbles, as we move through the annealing schedule. The optimum
search for $u$ was started with the cost function far away from zero, the
starting value being close to 0.1; the starting guess value of $u$ chosen here is
arbitrary, which could easily have been some other value far from the actual solution.
In our simulation the annealing schedule was a 2\% decrease in $T_{at}$, i.e.,
starting with an initial annealing temperature of $T_{at}=100$ after each
simulation step (or temperature step) $T_{at}$ was decreased by 2\% of its current value.
For each temperature step $T_{at}$ in the SA run, a maximum of 30 samplings
were carried out, however, in cases when 20 steps were successful from the
Metropolis sampling, the simulation was started at the next lower annealing
temperature obtained following the annealing schedule. In Fig.~\ref{uprofile}
we plot the profile of $u$ against the inverse of the annealing temperature
$T_{at}$. The convergence is rapid towards the correct value $u=0.6$, reaching
$u=0.5997$ after about 10 iterations. The simulation is carried on for more
steps to ensure that the obtained value corresponds to a final plateau.

During sampling for moves which pass the Metropolis test, the instantaneous values
of $u$ can show fluctuations at a particular annealing temperature 
(as one samples the system a number of times at a given annealing temperature)
and more so when 
the annealing temperature is high. This fluctuations in $u$ will decrease as we move
closer to the solution which is achieved at a lower annealing temperature. This is because
the Metropolis sampling scheme is such that lesser and lesser number of moves are 
accepted as the annealing temperature goes on decreasing. The histogram in the inset 
of Fig.~\ref{uprofile} depicts this fluctuations in $u$ for the entire range of annealing 
temperature in the SA sampling performed. Initially, the $u$ values show higher fluctuations
as these correspond to sampling at a higher annealing temperature, but these fluctuations
gradually decrease and near the vicinity of the solution $u$ values other than those close
to the actual solution are not accepted. Hence, the maximum normalized count in
the inset of Fig.~\ref{uprofile} correspond to $u=0.6$, which is the actual solution.

In Fig.~\ref{prob_dist} we plot the evolution of the bubble size distribution
$P_{T_{at}}(m)$ for some different annealing temperatures $T_{at}$. It can be
seen that at higher $T_{at}$ finer details of the population of the free
energy landscape, corresponding to the occupation of large bubble states
(high $m$) are not sampled, and consequently the probability distribution
falls off too quickly. At successively lower temperatures the profile found
from SA converges towards the true equilibrium distribution of the bubble.
While the curve for $T_{at}=100$ is far away from the equilibrium data,
the curve for $T_{at}=17$ exactly superimposes on the equilibrium profile.
The other two curves for $T_{at}=67$ and $T_{at}=47$ show the
intermediate 
dynamics and approach towards the equilibrium profile, respectively.
This is done so that we can follow the subtle features of the dynamics
and exactly track whether larger $m$-th sites get populated or not.
The observation is that as we proceed a
significant population develops at larger $m$ values with the decay from
higher $m$ to lower $m$ becoming more gradual. This is expected physically
since rare events need a finer sampling resolution (note the logarithmic
axis in Fig.~\ref{prob_dist}).

\begin{figure}
\includegraphics[width=0.8\columnwidth,angle=-90,bb=48 62 590 792]{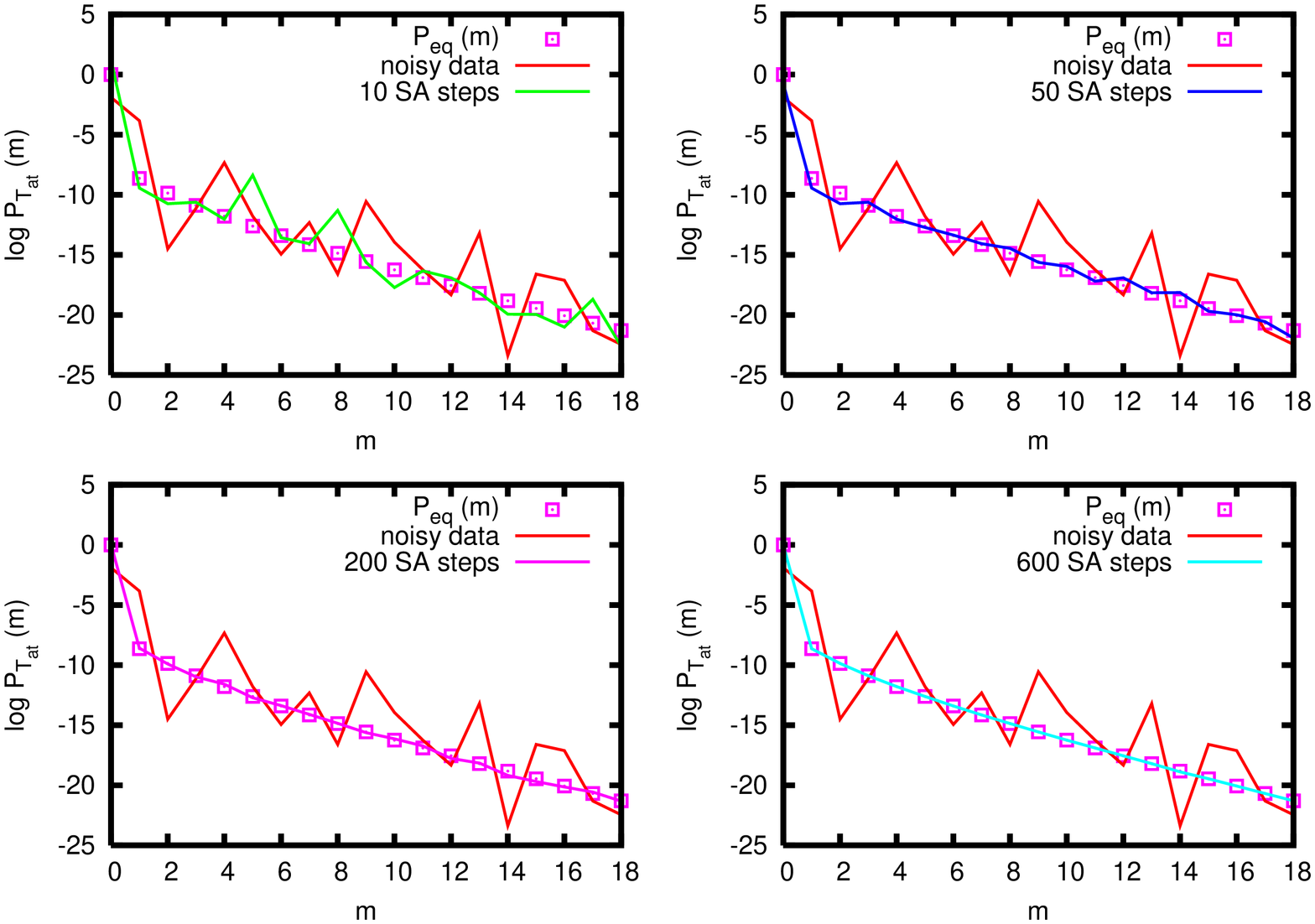}
\caption{(color online) Plot of log $P_{T_{at}}(m)$ against $m$ at various SA steps.The plot
for 600 steps matches the theoretical distribution.}
\label{noise}
\end{figure}

To test the robustness of the SA procedure, we also carried out calculations
to see how a very noisy distribution profile, created by introducing
reasonable perturbations onto the equilibrium distribution data
($P_{eq} (m) \pm \lambda \times rn$, $\lambda \sim 5$, $rn \in [0,1]$),
converges to the correct one. The results of the calculation are shown in
Fig.~\ref{noise}. 
The SA run was started with the zigzag profile (solid line) denoted as noisy data. 
The starting temperature was kept pretty large (at nearly $10^{6}$), since 
the cost function was of a large value to start with. In this case
the cost function can be written as
\begin{eqnarray}\label{eq3}
{\rm cost}_{(u=0.6)} = \sum_{i=1}^{m} ( P_{eq}(m)- P_{T_{at}}(m))^2 ,
\end{eqnarray}
which looks similar to Eq.(1), with the only interpretational change 
being that we are trying to find out the correct distribution, keeping the 
activation energy $u=0.6$ in all the cases. The goal of the SA run is to smoothen 
the noisy profile and reach the optimum distribution (given by symbols) for $u=0.6$. 
Gradually with the progress of the calculation at successive steps 
($10$, $50$, $200$ and $600$)
the profile becomes more regular, with the noisy structures vanishing and at
the $600$-th step a profile which matches the exact theoretical one is
generated as is evident from the Fig.~\ref{noise}.

\section{Conclusion}

Potential application of the SA algorithm presented in this work will be
analyzing experimental single molecule data. However, except
in one case there is no available experimental data.
The only experimental data that is available come from fluorescence
correlation measurement \cite{oleg}.  However, it is only a question of 
time until new experimental data become available. Considering this
the present communication tries to establish a ground work which
can be further extended in analyzing experimental data in future.

Another potential application of SA would be the subject of recent publication
\cite{walker} in which a novel setup for obtaining more accurate data for 
DNA denaturation has been proposed.
In particular, this setup has the potential to reveal position-resolved
data. Having these potential experiments in mind we wanted to find out the
potential of SA to actually tackle the involved analysis.
In the submitted work we show that this is actually possible.

As we show in the present work,
the SA algorithm is extremely insensitive to noisy input
data; even after adding random values of significant amplitude to the
input data the correct parameters are found. This was not necessarily
expected a priori, and this is what we consider the major result of our work.

In the recent work
we have shown how optimum parameters in systems of biological interest can
be found out with the use of stochastic optimization techniques. These results
have inspired us to look into more complicated problems, like a two dimensional
model in which not only the size of the bubble but the exact position of
cleavage of base pairs can be found out \cite{tobias}. 
In addition to that, the experimental work involving single molecule techniques
so far concentrates on homopolymer DNA. It will be interesting to see what
power SA exhibits when longer and heteropolymer structures are tested.
Work in this direction is in progress, which we wish to address in  
our future communications.

\ack
PC wishes to thank The Centre for Research on Nano Science and Nano Technology,
University of Calcutta for a research grant [Conv/002/Nano RAC (2008)].
SKB acknowledges support from Bose Institute through a initial start up fund.

\end{document}